%% file: InterMCore.tex
\documentclass[ERTS, 2026]{ERTS} 

\usepackage{graphicx}
\usepackage{amsmath}
\usepackage{amssymb}
\usepackage{xspace}
\usepackage{algorithm}
\usepackage{algpseudocode}
\usepackage{tikz}
\usepackage{url}

\usepackage{listings}
\usepackage{xcolor}
\usetikzlibrary{shapes, arrows, positioning, chains, calc}
\usetikzlibrary{arrows.meta}

\newcommand{\phylog}{{\sc phylog}\xspace}
\newcommand{\amcMCP}{AMC20-193\xspace}
\newcommand{\multicore}{multi-core\xspace}

\makeatletter
\AddToHook{cmd/maketitle/before}{\let\thetitle\@title}
\makeatother

\usepackage{hyperref}

\begin{document}

% Paper Title
\title{Interferences within a certifiable design methodology for high-performance multi-core platforms}

% Authors List
\author{%
        % Just reordre as you want
        Mohamed Amine KHELASSI\authorNumber{1},
        Felix SUCHERT\authorNumber{3},
        Abderaouf AMALOU\authorNumber{2},
        Benjamin LESAGE\authorNumber{4},
        Anika CHRISTMANN\authorNumber{5},
        Robin HAPKA\authorNumber{5},
        Jeronimo CASTRILLON\authorNumber{3},
        Mihail ASAVOAE\authorNumber{1},
        Mathieu JAN\authorNumber{1},
        Claire PAGETTI\authorNumber{4},
        Selma SAIDI\authorNumber{5}     
}

% Author Affiliations
\address{%
	\affiliation{1}{Université Paris-Saclay, CEA, List, F-91120, Palaiseau, France}{%
	}
	\affiliation{2}{Nantes Université, École Centrale Nantes, LS2N, CNRS UMR 6004, F-44000 Nantes, France}{%
	}    
	\affiliation{3}{Technische Universität Dresden, Dresden, Germany}{%
	}
	\affiliation{4}{ONERA, Toulouse, France}{%
	}
	\affiliation{5}{Technische Universität Braunschweig, Germany}{%
	}
}

% Create the title
\maketitle

\chead{\thetitle}

\pagestyle{fancy}

\thispagestyle{plain}

% License Information, provide first author's name "FirstName LastName".
% If more than one author, please add '~et al.', e.g. \licenseFootnote{John Doe~et al.}
\licenseFootnote{Mohamed Amine Khelassi~et al.}

% Abstract
\begin{abstract}%   %NOTE: Deleting the percentage after "{abstract}" may be lead to an extra leading space in the first line of the abstract, and this should be prevented.

The adoption of high-performance multi-core platforms in avionics and automotive systems introduces significant challenges in ensuring predictable execution, primarily due to shared resource interferences. Many existing approaches study interference from a single angle—for example, through hardware-level analysis or by monitoring software execution. However, no single abstraction level is sufficient on its own. Hardware behavior, program structure, and system configuration all interact, and a complete view is needed to understand where interferences come from and how to reduce them.
In this paper, we present a methodology that brings together several tools that operate at different abstraction levels. At the lowest level, PHYLOG provides a formal model of the hardware and identifies possible interference channels using micro-architectural transactions. At the program level, machine learning analysis locates the exact parts of the code that are most sensitive to shared-resource contention. At the compilation level, MLIR-based transformations use this information to reshape memory access patterns and reduce pressure on shared resources. Finally, at the system level, Linux cgroups enforce static execution constraints to prevent highly interfering tasks from running together.
The goal of our approach is to reduce memory interference and improve the system's  predictability, thereby easing the certification process of multi-core systems in safety-critical domains.

%The adoption of high-performance multi-core platforms in avionics and automotive systems introduces significant challenges in ensuring predictable execution, primarily due to shared resource interferences. In this work, we propose a methodology focused on identifying and mitigating such interferences through an interference-centric approach. This includes formal micro-architectural modeling (PHYLOG), machine learning based interference identification, code transformation using the MLIR framework, and static system configuration with Linux cgroups. 
%The goal of our approach is to reduce memory interference and improve the system's  predictability, thereby easing the certification process of multi-core systems in safety-critical domains.


\end{abstract}

\input{1-Introduction}
\input{3-Overview}
\input{4-Use_Cases_Evaluation}
\input{5-Methodology_Results}
\input{2-Related_work}

\input{6-Conclusion}

\section*{Acknowledgement}
This work is partially funded by the Deutsche Forschungsgemeinschaft (DFG, German Research Foundation) through the InterMCore project (Grant No. 505744711) and by the Agence Nationale de la Recherche (ANR) under Grant No. ANR-22-CE92-0066.

% ---------------------------------------------------------------------------------
\bibliographystyle{plainnat}
\PHMbibliography{erts}
% ---------------------------------------------------------------------------------

\end{document}

%% file: 1-Introduction.tex
\section{Introduction}%
\label{sec:introduction}

%\textcolor{red}{
%\begin{itemize}
%    \item Context: safety‑critical systems and multi‑core challenges
%    \item Motivation: interferences across CPU, memory, accelerators affect predictability
%    \item Problem
%    \item Contribution : “We propose a combined hardware/software methodology to model, analyze and reduce interferences.”
%    \item Paper structure paragraph
%\end{itemize}
%}

% The rapid evolution and growing complexity of high-performance \multicore platforms pose significant challenges in safety-critical systems, particularly within the automotive and avionics domains. These systems must comply with stringent certification standards such as ISO26262~\cite{ISO26262} in automotive and DO-178C/CAST-32A~\cite{CAST32A, DO178C} in avionics, which emphasize the need for rigorous timing predictability and efficient resource management. Heterogeneous \multicore architectures  exacerbate these challenges due to their complex interactions among mixed-criticality applications and diverse execution models.

The increasing integration of high-performance \multicore platforms in safety-critical domains such as avionics and automotive systems presents both opportunities and challenges. These platforms offer the computational capabilities required for modern workloads but make it difficult to meet strict timing and safety requirements imposed by standards like ISO 26262 and DO-178C \cite{ISO26262,CAST32A,DO178C}. A key source of difficulty lies in the complex and often unpredictable interactions between concurrent applications (or tasks within an application) competing for shared hardware resources.
These interactions lead to hardly predictable delays, that are called \emph{interferences}.

Understanding and controlling inter-core interferences is critical to ensure predictable system behavior. However, such interferences are difficult to model, analyze, and mitigate due to their dependency on hardware architecture, executive layer (e.g. operating system or hypervisor) and execution patterns. Without a structured methodology, this unpredictability complicates certification efforts and increases system design costs.

There has been extensive research to identify and analyze interferences.
However, each approach focuses on individual aspects 
such as timing anomaly analysis or \emph{micro-architectural} interference identification. 
By \emph{micro-architectural}, we mean abstracting the application and executive layer as a set of micro-transactions (e.g. core reading a data in a DDR bank).
Such methods, although valuable, do not provide the comprehensive view required to accurately model and mitigate interference across the complex interactions between software execution and hardware behavior. 
Therefore, there exists a clear demand for an integrated methodology that bridges hardware and software analyses, enabling more accurate interference modeling and mitigation.

\begin{figure}[hbt]
    \centering
    \includegraphics[width=1.05\linewidth]{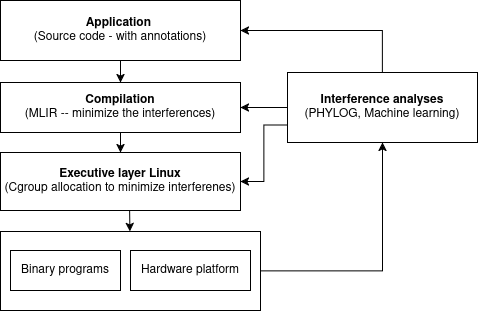}
    \caption{Overview of the interference-aware design methodology.}
    \label{fig:design}
\end{figure}

In this paper, we propose a combined hardware/software methodology for modeling, analyzing, and reducing interferences on \multicore systems. 
Our approach integrates several layers (see Figure \ref{fig:design}). We identify \emph{micro-architectural} interferences with \phylog approach, that relies on a platform model and targeted micro-benchmarks. 
We use machine learning to find code regions sensitive to those interferences and annotate the application code accordingly.
Thanks to this information, MLIR compiler optimizations are applied in order to reshape memory access to reduce contention and we configure Linux cgroups to control how resources are shared between tasks. 
%We also explore timing diversity as a safety layer a means to mask rare outliers caused by unpredictable execution delays.

% The InterMCore ANR project addresses this critical gap by developing an interference-centric design methodology that integrates formal modeling techniques with detailed benchmarking strategies. By capturing low-level timing behaviors through micro-architectural modeling, this approach facilitates precise system-level analyses, enabling robust identification of both direct and indirect interference effects. Additionally, the InterMCore methodology defines explicit software synthesis guidelines and establishes a unified compilation framework to systematically embed timing semantics and constraints into software development. This comprehensive integration of hardware modeling, formal analysis, and programming practices significantly enhances the predictability, reliability, and efficiency of safety-critical systems on complex \multicore platforms.

The remainder of this paper is structured as follows. The next Section~\ref{sec:methodology} details our methodology. Section ~\ref{sec:Use-Cases} presents our potential use cases. Finally, the conclusion~\ref{sec:conconlusion} outlines the directions for future research.

% Section ~\ref{sec:related} discusses related work

%% file: 3-Overview.tex
\section{Methodology}
\label{sec:methodology}

We aim to present an end-to-end methodology for detecting, quantifying, and reducing memory interference in multicore systems. To address memory interference in a practical way, we propose a flow that combines analysis, transformation, and runtime mitigation. The intuition is to integrate compile-time and runtime strategies guided by \phylog and machine learning analysis to both understand and control timing variability. To address the challenge, our methodology uses three complementary techniques: (1) interference analysis using the \phylog framework and machine learning, (2) compilation-level optimizations based on MLIR, and (3) %dynamic 
static resource allocation using Linux cgroups.

\subsection{Micro‑architectural Modeling}
\label{subsec:micro_arch_modeling}

% \textcolor{red}{
% \begin{itemize}
%         \item Build low‑level benchmarks to measure cache, bus, memory and accelerator interferences
%         \item Phylog
%         \item Interference analysis
% \end{itemize}}

The \phylog~\cite{phylog} methodology was designed to assist applicants in certifying their \multicore processors in the aeronautical domain. 
A platform description, in the \phylog modeling language~(PML), captures the knowledge about the characteristics of the platform based on the available documents and the applicant’s assessments. It also captures the target configuration, including hardware and software settings such as the mapping of applications hosted on the platform to cores.

The PML model serves as the backbone for the identification of \emph{interference channels} through interference calculus.
Interference channels, as per \amcMCP~\cite{amc20193}, are hardware resources whose use by an application might cause variations in its functional or temporal behaviour over its behaviour in isolation.
Interference calculus thus provides transactions combinations expected to be free from (or to be suffering from) interference.
Each transaction captures a path an application may exercise through hardware resources.

The interfering or interference-free combinations then support the validation of the model, understanding latent interference sources, and the quantification of said interference where applicable.
This is achieved through micro-benchmarks each exercising individual transactions.
Combinations of micro-benchmarks~\cite{microbench_hopscotch, courtaud2019improving}, matching the combinations of interest identified through interference calculus, should confirm the verdict of the analysis.
Otherwise, refinements to the model are required.

The application and executive layer are abstracted as a set of micro-architectural transactions. The identified interferences are also expressed as microarchitectural interferences.
Linking this knowledge to application code and behaviour is far from simple.
This is the reason why it is then combined with a complementary analysis explained below.

\subsection{Interference Modeling using Machine Learning}
\label{subsec:microbench}
We build on the microbenchmarking approach proposed by Courtaud et al~\cite{courtaud2019improving} to analyze how \multicore memory interferences affect code execution performance. Specifically, we gather some metrics from the Performance Monitoring Unit (PMU) and combine it with observed contention metrics (defined as the ratio between the execution time under contention and in isolation) we formulate an inverse machine learning problem~\cite{inverseML}. Our goal is to identify precisely which segments of the code are most vulnerable to performance degradation when competing for shared resources such as L2 caches, memory buses, and DRAM utilization. This approach offers a new way to pinpoint the code regions most sensitive to resource contention and create what we will call a code heatmap artefact that can be used as cost function for MLIR optimization pass and as input data for the cgroup allocation heuristic.

\subsection{Interference reduction via compilation}
\label{subsec:inter_reduc}

This part of the methodology focuses on reducing interference through compile-time and run-time techniques applied to parallel dataflow applications. Traditional compiler optimizations improve cache locality for single-threaded code, but they do not address the interference caused by parallel execution, especially when many dataflow nodes compete for a limited number of hardware threads. This leads to frequent task switching and extra cache misses.

To handle this, we first analyze the dataflow graph of the application and fuse compatible nodes to reduce the overall number of parallel tasks. This is done using an MLIR-based compilation flow (Etna), which transforms the application into a dedicated dataflow IR and applies fusion when nodes operate at the same rate and communicate unconditionally. Fewer tasks mean less oversubscription, fewer context switches, and lower memory interference.

At runtime, we complement this with the HARP resource manager, which monitors performance metrics and adjusts thread allocation. By prioritizing nodes that suffer more cache misses, HARP helps limit interference during execution without requiring changes to user code.

Together, these compile-time graph transformations and runtime resource adjustments reduce memory contention in parallel applications and improve execution predictability.

\subsection{Interference reduction via 
cgroups}
\label{subsec:inter_reduction_cgroups}

Cgroups provide a way to partition hardware resources in Linux, making it possible to restrict how tasks share CPU cores, memory, and other components. In our methodology, cgroups are used at runtime to prevent tasks that strongly interfere with each other from running at the same time.

We rely on an interference matrix, built from PHYLOG analysis and machine learning results, which quantifies how much slowdown each pair of tasks causes when they run together. Based on this matrix, we build a static cgroup hierarchy in which tasks that show high mutual interference are separated into different cgroups and controlled using the freezer subsystem, ensuring they do not execute concurrently. Tasks that do not interfere significantly can be grouped together.

This static cgroup configuration helps enforce predictable behaviour by isolating interfering tasks at the operating-system level, complementing the compile-time and program-level optimizations used in other parts of the methodology.

\section{How things interact}

Our methodology integrates distinct yet complementary layers to reduce memory interference in interference-aware design flow.

\begin{itemize}
    \item PHYLOG performs a platform-level analysis based on a formal model of the hardware and software configuration. It identifies potential interference channels using micro-architectural transactions and interference calculus.
    \item Machine learning analysis processes data from targeted microbenchmarks and performance counters. It pinpoints which code segments are most sensitive to interference, producing a code heatmap that characterizes application-level interference behaviour.
    \item MLIR-based transformations use this heatmap as a cost model to reshape memory access behaviour
    \item The cgroup configuration heuristic takes as input both the PHYLOG interference channels and the machine learning-based heatmap. It uses this joint interference characterization to assign tasks to cgroups and tune resource limits.
    % \item Timing diversity acts independently of the other steps but complements them by ensuring that rare execution time outliers do not compromise system correctness. It uses redundant execution and deadline monitoring to mask anomalies
\end{itemize}

%% file: 4-Use_Cases_Evaluation.tex
\section{Use-Cases \& Evaluation}
\label{sec:Use-Cases}

In our evaluation, we focus exclusively on the Raspberry~Pi~4 platform, which features four Cortex--A72 cores with private 32\,KB L1 data caches and a shared 1\,MB L2 cache. The system runs Raspbian, providing a lightweight environment suitable for rapid instrumentation. As the test workload, we use a parameterized matrix multiplication kernel, where the sizes $N$, $M$, and $K$ control the dimensions of matrices~$A$ and~$B$. This setup allows us to study how variations in memory-access patterns affect interference sensitivity and prediction accuracy.

\begin{lstlisting}[language=C]
#define N 256
#define K 4096 
#define M 9 

int main() {
    int A[N][K];
    int B[K][M];
    Initialize_Matrix_Random(N, K, A, 1, 10);
    Initialize_Matrix_Random(K, M, B, 1, 10);
    int C[N][M]; 
    MM(A,B,C); 
    return 0;
}
\end{lstlisting}

To study the sensitivity to interferences we measured the number of L2-cache accesses by fixing one of the parameters 
$M$, $N$, or $K$ to 32 while varying the remaining parameter. 
The varying parameter started at 64 and increased with a step of 128 until it 
exceeded 4096. 
The resulting curve, shown in the figure, indicates that matrix multiplications 
with a large value of $K$---corresponding to irregular matrix shapes---produce 
significantly more L2-cache accesses. 
Although such large $K$ values may appear unusual, they are common in modern 
inference workloads, particularly in Transformer-based models and LLMs, where 
many matrix operations involve highly non-square shapes (with $K$ much larger 
than $M$ and $N$). 
Moreover, a higher number of L2 accesses is advantageous for studying 
interference effects.

\begin{figure}[t]
    \centering
    \includegraphics[width=\linewidth]{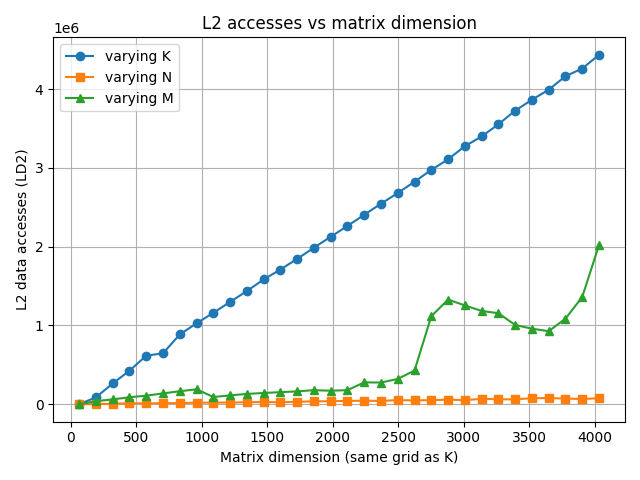}
    \caption{Sensitivity of L2 Accesses to Variations in Matrix Dimensions (M, N and K) starting from 64 with a step of 128, the other dimensions are fixed into 32.}
    \label{fig:placeholder}
\end{figure}

%% file: 5-Methodology_Results.tex
\subsection{Micro-architectural Modeling}

\begin{figure*}[t]
    \centering
    \includegraphics[width=0.8\linewidth]{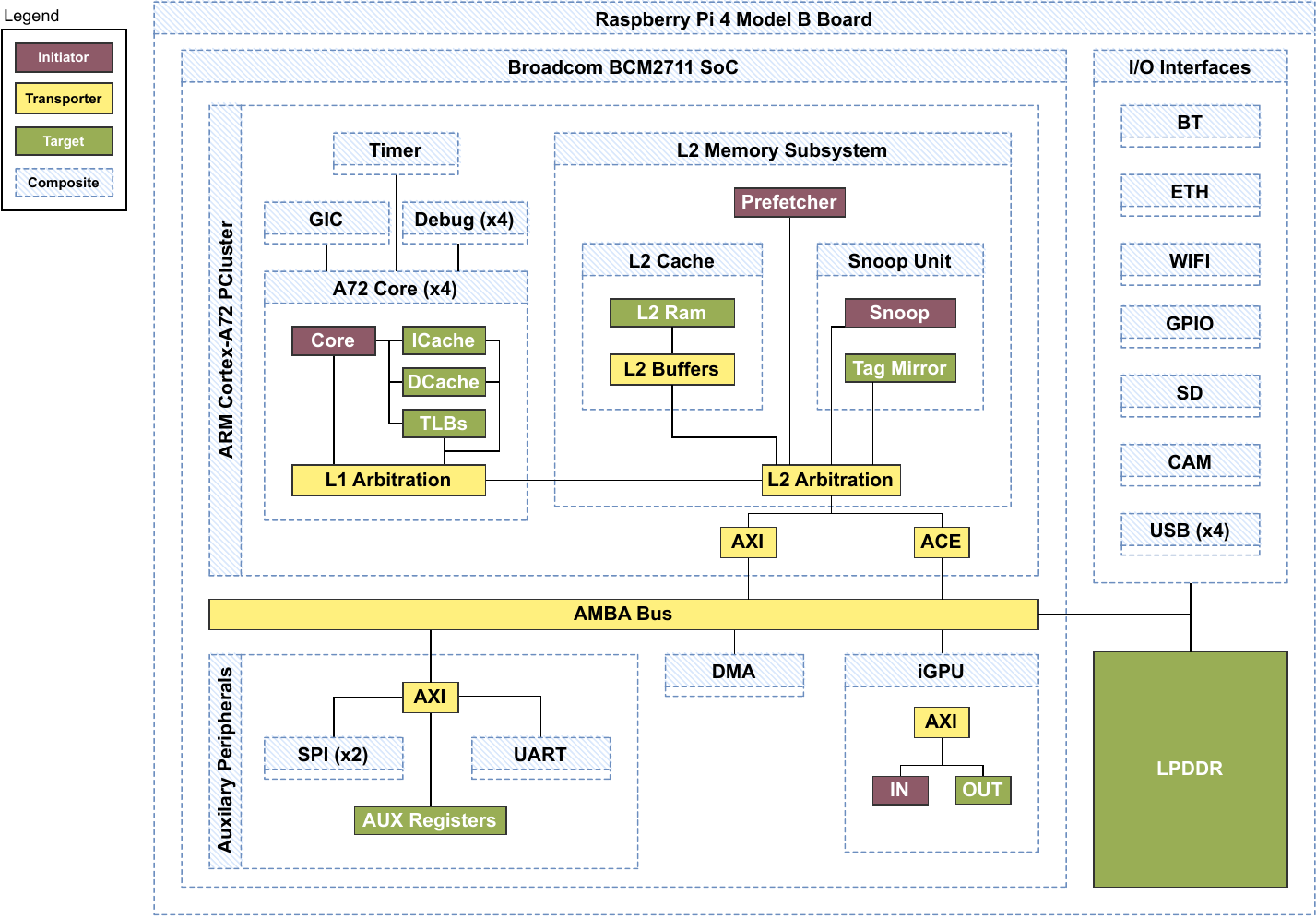}
    \caption{Overview of the PML model for the Raspberry Pi 4 Board}
    \label{fig:rpi4_pml}
\end{figure*}

To support the identification of potential interference channels on the target Raspberry Pi 4 platform, we built a model of the board. Figure~\ref{fig:rpi4_pml} presents an overview of the identified hardware resources. Hardware resources have been classified (and color-coded) following the PML nomenclature into Initiators, Transporters, or Targets for transactions.
The model was built through a first review of the available documentation on the board~\cite{trm:rpi4}, the embedded Broadcom BCM2711 System-on-Chip~(SoC)~\cite{trm:bcm2711}, and its ARM A72 processor~\cite{trm:a72}.

The Raspberry Pi 4 board comprises a Broadcom BCM2711 SoC, configurable LPDDR4 SDRAM, and a number of supporting interfaces such as USB, GPIO, HDMI, etc. Our efforts focused on capturing the resources relevant to considered applicative use case; the model only captures interfaces and peripherals at the board- and SoC-level with a high granularity. 
We consider a conservative model, exemplified by the integrated VideoCore GPU (\emph{iGPU}), where each such device is a composite with a single target and a single initiator, sharing a common port.
This choice is further reinforced by the lack of documentation regarding those devices. 
Revisions will be required under a different use case, or if these peripherals indeed affect the considered applications.

The ARM Cortex-A72 Cluster on the BCM2711 does feature 4 cores, each with its own Level 1 (L1) caches for instruction (ICache), data (DCache), and page descriptors (TLBs). The L2 memory subsystem is shared between the cores, featuring a unified L2 Cache, a Snoop Control Unit for memory coherency, and the L2 prefetcher. Note that while there are two levels of TLB per core, we elect to conservatively model them as a single target. The TLBs act as single entry block; the only path to the L2 is through the L1, and the only path from the L1 goes through the L2~\cite{pml:example1}.

The interference analysis, without any additional mitigation from the hardware or software configurations, highlights the shared resources as sources of interference: the L2 subsystem, the AMBA Bus, and the LPDDR.
The analysis also flags private caches as a source of interference. The L1 and L2 caches are inclusive, such that a line in the L1 must also be present in the L2. An eviction from the L2 may thus result in the same line being evicted from the L1. Without any additional restrictions, a private L1 may thus be impacted by other cores, or the L2 prefetch.
Similarly memory coherency, enforced by the snoop, may result in the invalidation of data in the L1 and L2 caches due to requests from other cores, or from devices through the coherence ACE and AXI ports.

\subsection{Interference Modeling using Machine Learning}
\label{sec:ml}
We build on the microbenchmarking method of Courtaud et al.~\cite{courtaud2019improving} to quantify how shared-memory pressure affects execution time. Our approach combines timing measurements, PMU events, and a dynamic extraction of the instruction trace. The goal is to identify the regions of code that suffer the highest slowdowns when sharing L2, buses, or DRAM. The final output is a \emph{code weights vector} usable as a cost function for MLIR optimizations and as input to cgroup allocation policies.

\subsubsection{Phase 1: Collecting Interference Timing}
\textbf{Instrumentation and execution setup.}  
We reuse the microbenchmarks of Courtaud et al.~\cite{courtaud2019improving} as training and validation victims. Each benchmark is executed twice: (i) isolation and (ii) concurrently with aggressors.  
Aggressors are hand-written stressors targeting specific microarchitectural components: one precisely loads on a specified LLC bank continuously (inspired from~\cite{cacheBank}); another saturates both the memory bus and DRAM. Each aggressor runs on a dedicated core in an infinite loop. We monitor their activity (aggressiveness) using hardware event counters (how much LLC misses we have, how much data are passing through the bus).
\textbf{Timing measurement.}  
Each victim contains explicit observation points inserted (by hand) before and after loop nests or simple loops. At each point, we record the execution time in isolation and under interference. We repeat every run 100 times in both conditions.
\textbf{Ground truth (delta).}  
For each observation point, we compute the slowdown:
\[
\Delta = \frac{T_{\text{interference}}}{T_{\text{isolation}}}.
\]
This ratio is the ground truth used to train our learning model.

\subsubsection{Phase 2: Collecting Instruction and Data Adresses About the Victim}
\textbf{Assembly trace extraction.}  
This phase is independent of timing measurements. Using \texttt{gdb}, we extract the full instruction trace of the victim (we represent a program by an execution trace): instruction address, instruction type, and the corresponding data address when applicable.
\textbf{Trace transformation.}  
The trace is converted into a structured sequence:
\[
\langle \text{InstrAddr},\; \text{InstrType},\; \text{DataAddr} \rangle.
\]
This heavy parsing step is performed only once per program.

The combined dataset (timing deltas + instruction-level features) forms the input to our machine-learning model. In addition to extracting the instruction trace, we also use the collected data addresses to determine which LLC bank the victim accesses most. The bank is inferred from the physical-address bits associated with the LLC slices. The dominant bank is then targeted by an aggressor configured to saturate the same cache region.

\subsubsection{Phase 3: Training}
We train a BERT-style Transformer model. The encoder is pretrained on assembly code to learn structural and semantic patterns, following the same pretraining philosophy as CAWET~\cite{cawet}. We fine-tune it with pairs $(\text{execution trace}, \Delta)$ using an RMLSE loss, which penalizes underestimation more than overestimation.

During fine-tuning, we observe poor generalization when the model relies only on the delta-regression objective (42\% accuracy).
To investigate this behaviour, we extract the last attention matrix of the Transformer for each inference. This matrix has size $N \times N$, where $N$ is the length of the trace. Most of the attention mass concentrates on instruction types that are not LOAD or STORE, nor on addresses. To align the model with actual memory behaviour, we replace timing by the number of L2 accesses. For each victim, we build an \texttt{L2\_Access\_MAP}, a vector counting the number of L2 accesses (using performance event) per region (these regions follow the observation-point boundaries). We also reduce the attention matrix into a vector by summing all elements of each row, then aggregating these values into the same regions, yielding \texttt{Attention\_MAP}.

Both vectors are normalized so that every element lies in $[0,1]$: \texttt{L2\_Access\_MAP} is scaled by the total number of L2 accesses, and \texttt{Attention\_MAP} by its maximum value. Their dot product produces a
correlation score bounded by the number of observation points $N_{\text{obs}}$. This score is added to the loss as an extra term, encouraging the model to align its attention with actual L2-access behaviour while predicting deltas.

Let
\[
C = \text{dot}(\text{Attention\_MAP}, \text{L2\_Access\_MAP}).
\]

A perfect alignment yields $C = N_{\text{obs}}$. 
We convert this into a penalty by taking the distance from the ideal value:

\[
\text{Penalty} = N_{\text{obs}} - C.
\]

The final loss is:

\[
\mathcal{L}
  = \text{RMLSE}(\Delta_{\text{pred}}, \Delta_{\text{obs}})
  + \lambda \cdot (N_{\text{obs}} - C),
\]
With RMLSE formula:
\[
\mathrm{RMLSE}(y,\hat{y}) =
\sqrt{
    \frac{1}{n}
    \sum_{i=1}^{n}
    \left(
        \log(1 + \hat{y}_i) - \log(1 + y_i)
    \right)^{2}
}
\]
where $\lambda$ controls the influence of this correlation-based term. In this paper we fix $\lambda$ to $0.5$. With this new loss we get a higher accuracy of 72\% (compared to 42\% we using just RMLSE).

\subsubsection{Results}

\begin{figure}[t]
    \centering
    \includegraphics[width=0.7\linewidth]{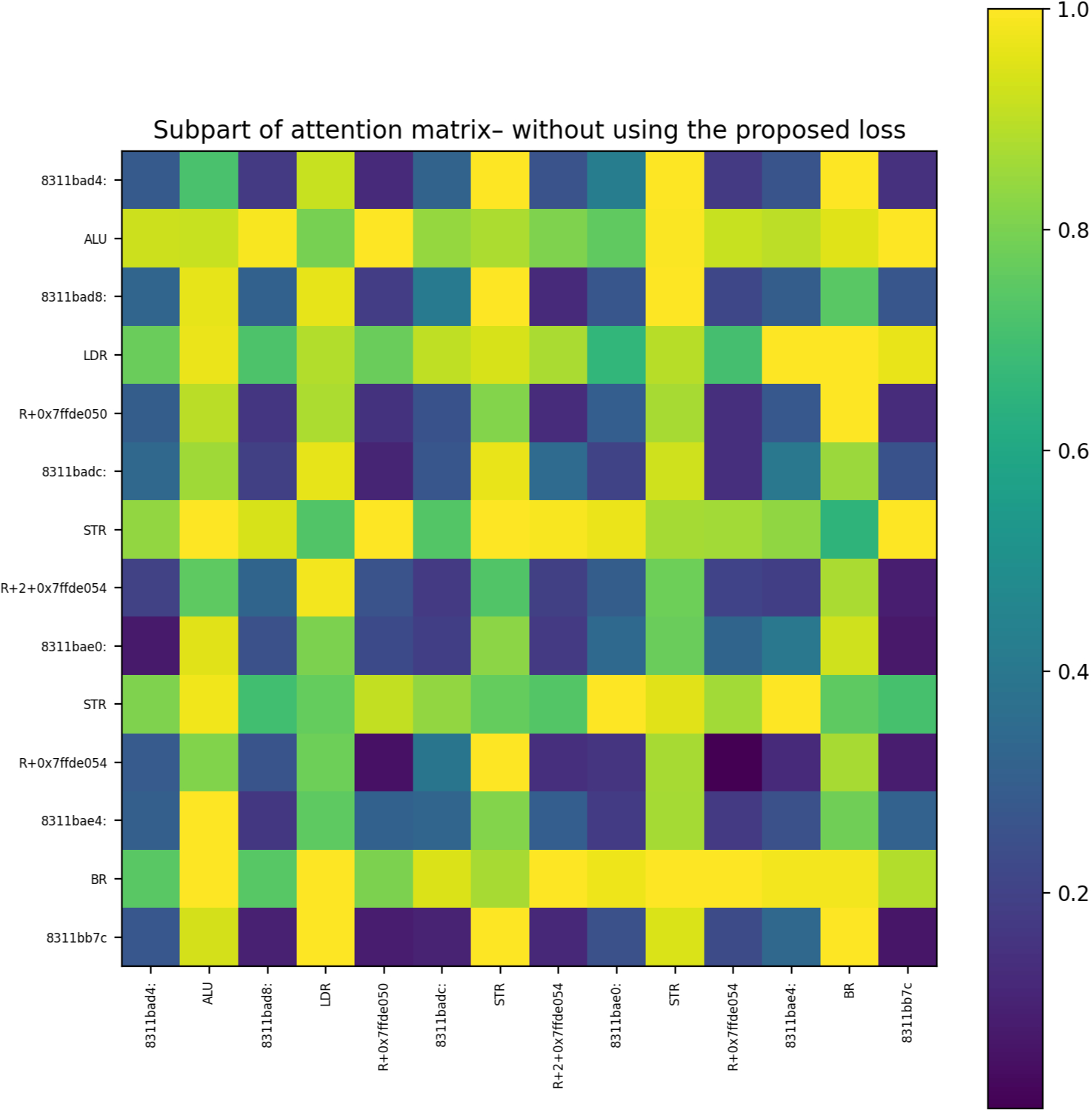}
    \includegraphics[width=0.7\linewidth]{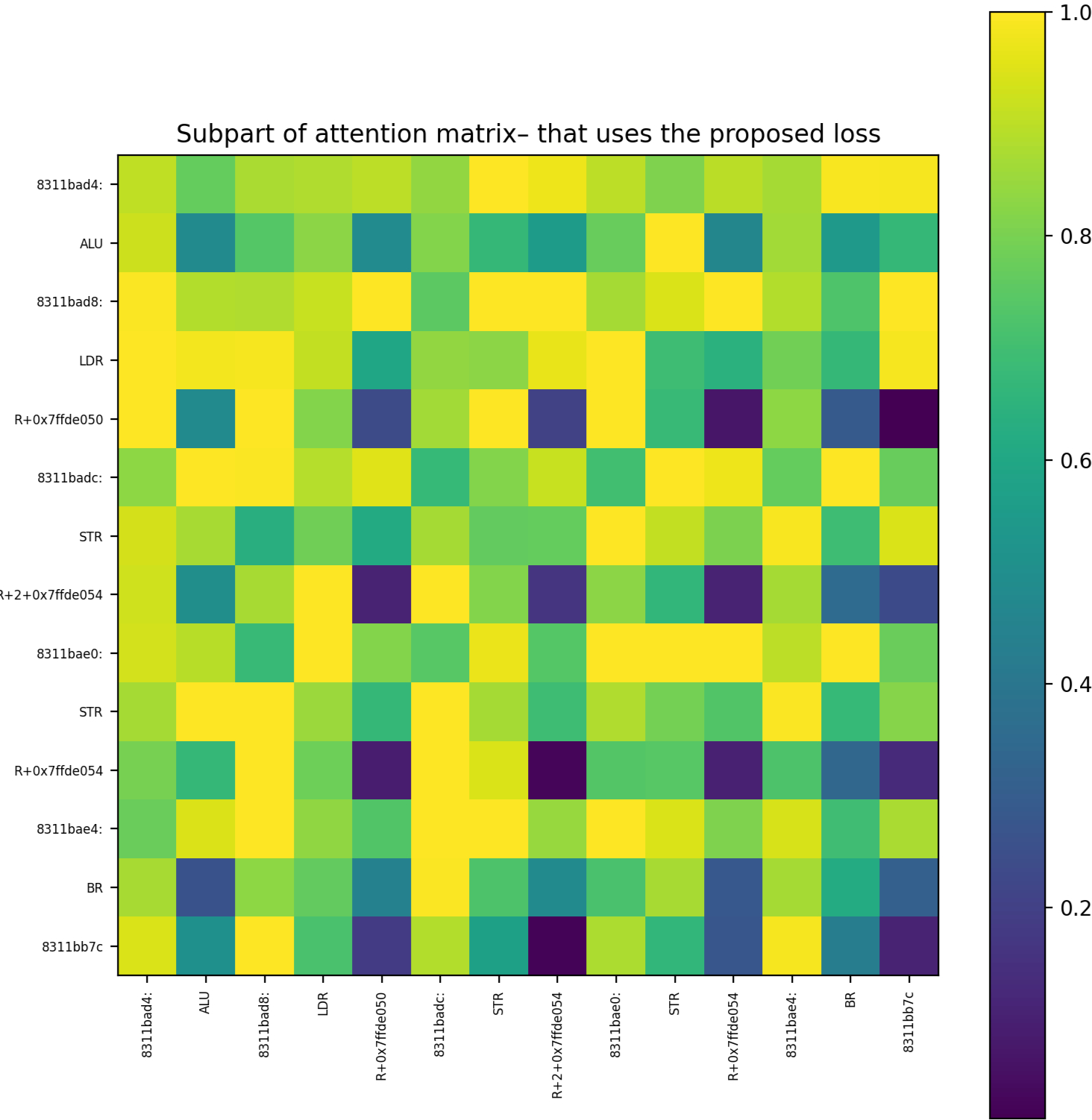}
    \caption{Final-layer attention matrices without (top) and with (bottom)
    the correlation loss.}
    \label{fig:attention_comparison}
\end{figure}

\paragraph{Attention Matrix: With and Without the Correlation-Based Loss.}
Figure~\ref{fig:attention_comparison} shows a subpart of the transformer’s final-layer attention matrix when trained (i) without the correlation-based penalty and (ii) with the proposed loss term. Without the penalty, the model concentrates most attention on instruction types, exhibiting little sensitivity to memory behaviour. With the penalty, attention shifts toward instruction regions that correspond to heavy L2 activity, indicating that the model better captures the actual sources of interference.

\paragraph{Accuracy Under Varying $K$.}
To assess how the model generalizes, we experiments on matrix multiplication, we vary the parameter $K$ in the matrix multiplication kernel while keeping $N$ and $M$ fixed. Increasing $K$ changes the stride and reuse distance of the matrix~$B$, directly affecting the L2 pressure and interference sensitivity.

\begin{figure}[t]
    \centering
    \includegraphics[width=\linewidth]{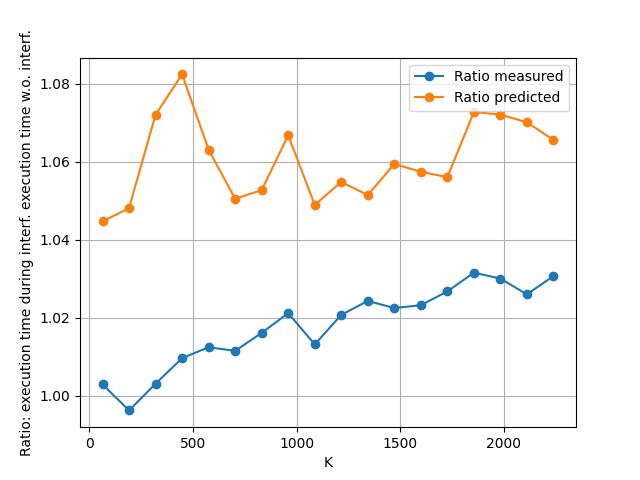}
    \caption{Prediction accuracy when varying parameter $K$ in the matrix
    multiplication kernel.}
    \label{fig:accuracy_k_variation}
\end{figure}

Figure~\ref{fig:accuracy_k_variation} reports the accuracy of the prediction for 
different values of $K$. We measure the ratio (Delta) when running under maximum interferences and when running the matrix multiplication without running our aggressors. We also give the average of the ratio estimated by our Transformer model. The results show that for matrix multiplication under variation of $K$, our estimations never underestimate the delta while keeping a reasonable distance from the measurements. This will help us to use the machine learning model during compilation as a cost function to indicate the sensitivity of the program to interferences.

At this stage, our approach operates exclusively on binary code, which limits its applicability to MLIR-based workflows. A key direction for future work is to extend the front-end to MLIR and its relevant dialects, and to train the Transformer directly on this intermediate representation so that interference patterns can be learned and mapped at the MLIR level rather than from low-level assembly.

\subsection{Interference reduction via compilation}

\begin{figure}[t]
    \centering
    \includegraphics[width=\linewidth]{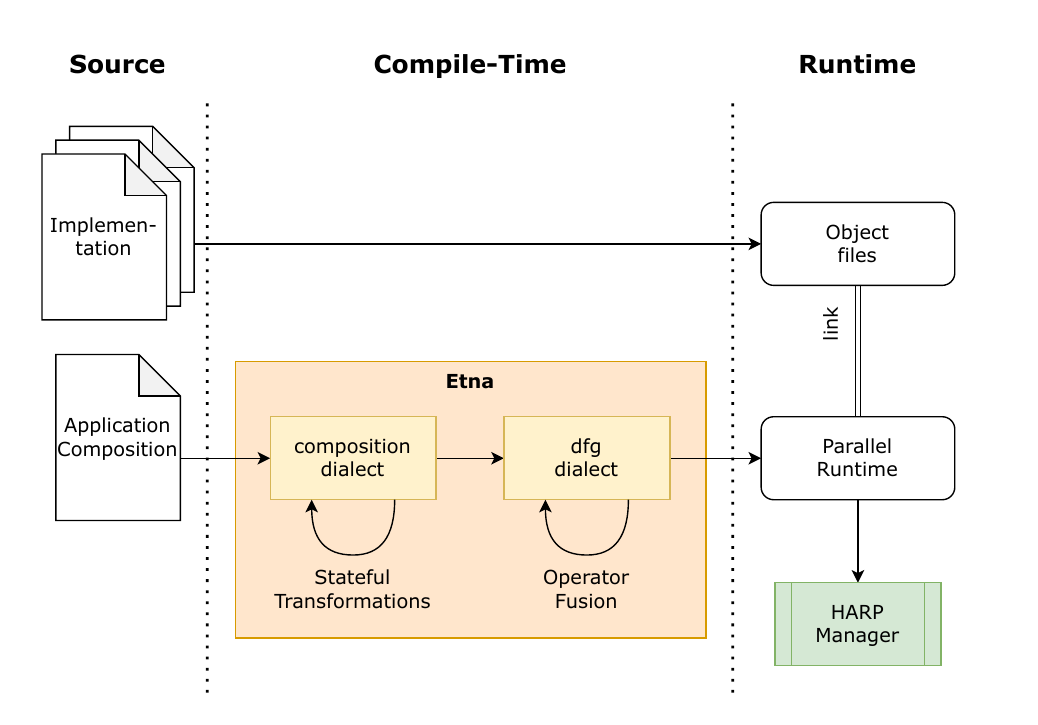}
    \caption{Overview of the compilation flow.}
    \label{fig:compileflow}
\end{figure}

An application running on a multi-core system may suffer from interference, primarily due to memory access contention with shared hardware resources.
% The biggest interference source in this scenario is memory usage.
A substantial amount of the memory interference potential within an application is determined at compile time.
% The execution schedule determines the order of memory accesses and thereby whether effects like cache locality can be exploited or data needs to be evicted and refetched -- the program interfering with itself.
Compile-time transformations to reduce cache misses and exploit data locality are good ways to reduce memory interference within a single-threaded application.
% TODO(Felix): add citations 
However, they fall short in the context of applications that exploit parallelism and have an inherent dataflow graph.
As part of our methodology, we aim to close this gap by providing means to reduce the interference within a parallel application on a dataflow graph level, using a combined compile time and run time approach.

\subsubsection{Interference Model}
\label{sec:compilation:model}

In parallel dataflow applications, task switching acts as an additional source of memory interference not caught by classic single-thread focused optimizations.

Dividing an application into several threads executing in parallel often leads to an over-saturation of the underlying hardware threads with work.
The result is task switching, which incurs a significant memory interference overhead, as the data gets loaded in and out of cache every time a switch occurs.
However, since in a normal execution environment, the operating system scheduler allocates resources, task switching itself is unpredictable and adapting your application to the constantly changing available resources would require significant changes to user code.
% solution: introducing an intermediate arbitrator/scheduler

Applications modeling a dataflow graph tend to be especially affected by the over-saturation problem.
Even simple computational graphs often have more nodes than the target hardware has threads.
However, these graphs offer the advantage of consisting of compartmentalized, separate tasks that communicate via a well-defined interface.
Yet, not all nodes in a graph have the same complexity and worst-case execution time.
Thus, fusing neighboring nodes of the graph can be a means to reduce memory interference by reducing the over-saturation of threads and eliminate overhead.

To address this problem, we propose a set of compile-time optimizations (\autoref{sec:compilation:opts}) to alleviate the over-saturation problem.
A run-time resource manager (\autoref{sec:compilation:rt}) complementary to our cgroups method (\autoref{sec:cgroups}) allows the application to intelligently adjust at runtime to changes in the available resources.
Our system is shown in \autoref{fig:compileflow}.

\subsubsection{Compile-Time Optimizations}
\label{sec:compilation:opts}

The core of our dataflow compilation flow is built atop \emph{Etna}~\cite{soldavini2024etna}, a compiler toolchain based on MLIR~\cite{mlir}.
% Its flow is described in Fig ??
It ingests the description of an application as a sequential composition of functions and derives a dataflow graph from it~\cite{suchert2023condrust}.
For this graph, a deterministically parallel runtime is generated, exploiting both pipeline and data parallelism.

Our compiler frontend processes an application composition expressed in a high-level syntax modeled after the programming language C.
It is used to compose the high-level algorithm of the application from functions implemented in other source files of the code base.
Besides function definition, declaration and calls, the syntax supports C-style loops and recursion.The \texttt{composition} MLIR dialect models this representation on an IR level.
It is used to understand the parallelism opportunities within an application by also modeling the mutability of data.
With this information, we can apply IR transformations that allow the exploitation of data parallelism throughout the application where possible.

This representation is translated into the \texttt{dfg} dialct, which models the graph using dedicated IR constructs for dataflow nodes and edges~\cite{soldavini2024etna, bi2024leveraging}.
The underlying idea of this model is, that every single node will be mapped to an OpenMP~\cite{dagum1998openmp} task.
As a result of that, the fusion of nodes at a \texttt{dfg} level will immediately lead to a reduction of task pressure.
Since the dialect offers a clean definition of the nodes and their communication patterns, two nodes can be fused in a safe and side-effect free fashion.
%TODO (felix): dfg code snippet?
Two nodes are eligible for fusion if they are executing at the same rate\footnote{In KPN networks, this would be labelled as a 1:1 connection.} and connected via a dataflow edge that is triggered unconditionally.
We use the model presented in \autoref{sec:ml} to identify nodes suitable for fusion.

At the end of the compilation process, we lower our \texttt{dfg} representation into the \texttt{llvm} dialect and link the generated object file against the other object files produced from the remaining code of the developer.
Using our compilation flow, we generate a deterministic parallel runtime, optimized for providing better performance\cite{suchert2023condrust}, while also reducing memory interference.
%TODO(felix): more emphasis on the determinism part??

\subsubsection{Run-Time Management}
\label{sec:compilation:rt}

In order to mitigate the memory interference caused by task switching at runtime, as well as the constantly changing availability of hardware resources, a method to influence the resource assignment within the application is necessary.
Depending on the available resources, nodes with higher demand (driven by computation time and cache usage) should be prioritized.

For this, we leverage HARP~\cite{khasanov_middleware25}, a Linux-integrated resource management approach.
It provides a unified resource allocation interface and jointly manages all application threads for an efficient resource utilization.
Its support for the OpenMP programming model means that we can leverage it without needing to change any user code.
Originally, this work was geared towards heterogeneous processors and improving energy efficiency.
% TODO(felix): HOW?
It analyzes on-line performance metrics of an application to adjust resource allocation decisions at runtime.
For this work, we introduce the metric of cache misses into the decision making process of resource assignment.
The more cache misses a thread shows due to task switching, the higher its priority in the resource assignment to avoid costly interference.

This runtime system is complementary to our cgroups approach, as it works exclusively on an application-thread level, rather than across processes.
It can adapt the application as a whole to the number of available threads and prioritize individual nodes with high interference potential.

\subsection{Interference reduction via Cgroups}
\label{sec:cgroups}
Linux control groups (cgroups) are a kernel feature that enables hierarchical partitioning of system resources, such as CPU, memory, and I/O, among groups of processes or tasks. They provide isolation and limit enforcement, making them suitable for mitigating interferences in multi-core environments by controlling resource allocation and task execution behaviour~\cite{cgroupv2}. In our methodology, cgroups serve as a runtime mechanism to complement compile-time optimizations (e.g., via MLIR), enforcing hardware-level constraints to reduce memory contention in safety-critical systems.
The context for this cgroup-based mitigation is the broader interference-aware design flow (Figure 1), where micro-architectural interferences identified by PHYLOG and machine learning (Sections 2.1 and 2.2) inform code transformations and resource configurations. Unlike prior applications of cgroups that focus on temporal isolation~\cite{chen2021schedguard, andriaccio2025scheduling} or frequency scaling~\cite{kim2018reducing}, our approach targets memory interferences by statically configuring cgroups to prevent concurrent execution of contending tasks, enhancing predictability for certification under standards like DO-178C.

\subsubsection{System Model and Assumptions}

We model the system as a set of tasks $\mathcal{T}
 = \{\tau_1, \tau_2, \dots, \tau_n\}$ deployed on a multi-core platform, where each $\tau_i$ is provided as a binary executable. Interferences are captured in a symmetric matrix $I \in \mathbb{R}^{n \times n}$, where entry $I_{ij}$ quantifies the degradation factor (e.g., slowdown in execution time or increase in cache-miss rate) when $\tau_i$ and $\tau_j$ run concurrently. A value $I_{ij} > 1$ indicates contention, with higher values signifying greater sensitivity (e.g., $I_{ij} = 2.5$ implies a 2.5$\times$ slowdown).
 
Assumptions include:

\begin{itemize}
    \item Availability of the interference matrix $I$, derived from upstream analyses (e.g., PHYLOG micro-benchmarks and machine learning heatmaps). In this work, we treat $I_{ij}$ as a scalar representing the worst-case degradation across the entire execution. This assumption can be refined in future work by replacing each $I_{ij}$ with a vector $I_{ij} = [I_{ij}^{(1)}, \dots, I_{ij}^{(k)}]$, where each component captures the interference between specific code regions of $\tau_i$ and $\tau_j$, enabling finer-grained mitigation strategies
    \item Support for cgroup v2 controllers: \texttt{cpu} (for scheduling weights and shares), \texttt{cpuset} (for CPU affinity and core pinning) and \texttt{freezer} (for pausing task execution)
    \item Static deployment: Configurations are determined pre-runtime
\end{itemize}

The goal is to generate a cgroup hierarchy that minimizes aggregate degradation by isolating high-interference pairs, thereby controlling task concurrency.

\subsubsection{Heuristic for Cgroup Deployment}

We employ a static heuristic to transform $I$ into a cgroup deployment plan. The algorithm identifies contention-sensitive pairs or clusters (where $I_{ij} > \theta$, with $\theta$ a user-defined threshold, e.g., 1.5) and assigns them to separate sub-cgroups under a root cgroup. This enables serialization via the \texttt{freezer} controller, while global limits at the root enforce overall constraints.

The pseudocode of the heuristic is as follows:

\begin{algorithm}[H]
\caption{Static Cgroup Hierarchy Construction}
\label{alg:cgroup}
\begin{algorithmic}[1]
\Require Taskset $\mathcal{T} = \{\tau_1,\dots,\tau_n\}$, interference matrix $I$, threshold $\theta > 1$
\Ensure No pair with $I_{ij} > \theta$ can run concurrently
\State All tasks initially reside in the root cgroup (default kernel behaviour)
\For{each unique task pair $\{\tau_i,\tau_j\}$ with $i < j$ and $I_{ij} > \theta$}
\State Create sub-cgroup $C_i$ (if not already existing) and move $\tau_i$ into $C_i$
\State Create sub-cgroup $C_j \neq C_i$ (if not already existing) and move $\tau_j$ into $C_j$
\State Enable the freezer controller on both $C_i$ and $C_j$
\EndFor
\State Tasks without any high-contention pair may remain in the root cgroup
\end{algorithmic}
\end{algorithm}

\subsubsection{Experimental Evaluation}
\label{sec:experiments}

To validate the effectiveness of our cgroup-based interference mitigation strategy, we conduct a series of experiments, the goal is to demonstrate that our static cgroup hierarchy and runtime freezer mechanism can effectively restore temporal predictability in the presence of heavy memory contention.

\paragraph{Experimental Setup:}
The experiments are performed on a Raspberry Pi 4 (Broadcom BCM2711, Quad-core Cortex-A72). The system runs Raspberry Pi OS Lite with Cgroup v2 enabled.
We utilize the RT-Bench framework~\cite{nicolella2022rt}, which extends the TACleBench suite for periodic real-time execution. Two tasks constitute our workload:
\begin{itemize}
    \item We use \texttt{matrix1} as a \textbf{victim task ($\tau_{victim}$)}, a $256 \times 4096 \times 9$ integer matrix multiplication program. Its $\approx$4MB working set exceeds L2 capacity, making it sensitive to cache evictions. It has a period of $T = 100$ ms and a relative deadline of $D = 30$ ms to simulate a tight real-time constraint.
    \item \textbf{Contender Task ($\tau_{noise}$):} We use the \texttt{bandwidth} benchmark from the IsolBench suite as the interference source. This synthetic benchmark issues sequential writes to a large buffer, saturating memory bandwidth. It has a period of $T = 200$ ms and a relative deadline of $D = 200$ ms.
\end{itemize}

To maximize resource contention, both the victim task and the interfering workload are pinned to different physical cores (which are also isolated from the OS execution) using the \texttt{cpuset} controller. This configuration forces the tasks to compete for CPU, L1/L2 caches, and memory bandwidth.

We evaluate our workload in three execution scenarios over 100 periodic jobs:

\begin{enumerate}
    \item \textbf{Solo:} $\tau_{victim}$ runs in isolation to establish a baseline worst-case execution time (WCET).
    \item \textbf{Interference (Unprotected):} $\tau_{victim}$ and $\tau_{noise}$ run concurrently in the same core without any mitigation.
    \item \textbf{Protected:} $\tau_{victim}$ and $\tau_{noise}$ are placed in separate cgroups inside the same core as per Algorithm~\ref{alg:cgroup}. A userspace monitor program executed in another core polls the CPU usage of the cgroups (at 50ms intervals) and freezes the $\tau_{noise}$ group when the CPU bandwidth consumption exceeds a safety threshold (30\%), effectively serializing access during peak contention.
\end{enumerate}

\paragraph{Results and Analysis:}

\begin{figure}[ht]
    \centering
    \includegraphics[width=\linewidth]{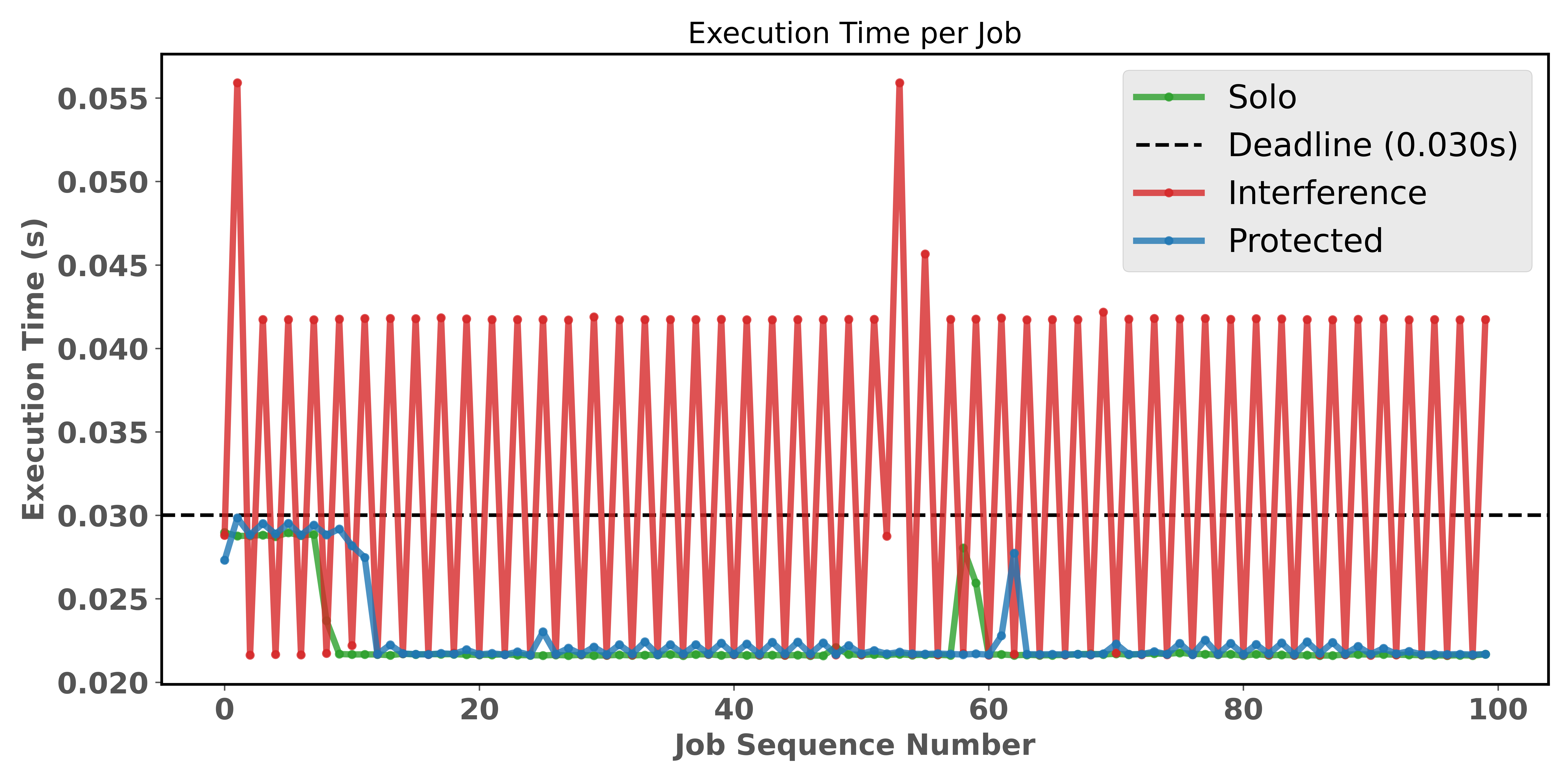}
    \caption{Job-wise execution times for \texttt{matrix1}. The dashed line represents the 30ms deadline}
    \label{fig:execution_times}
\end{figure}

\begin{figure}[ht]
    \centering
    \includegraphics[width=\linewidth]{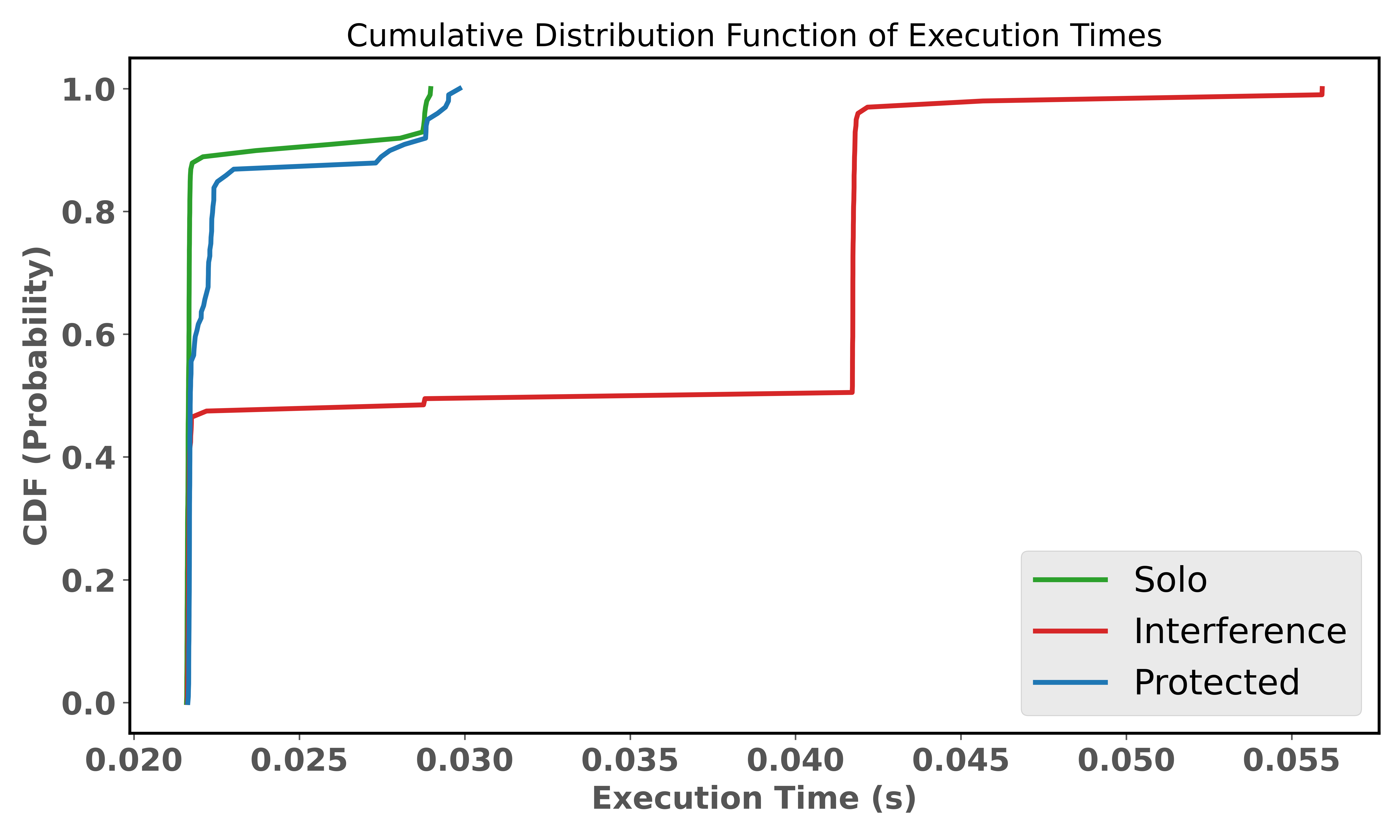}
    \caption{Cumulative Distribution Function (CDF) of execution times.}
    \label{fig:cdf}
\end{figure}

\begin{figure}[ht]
    \centering
    \includegraphics[width=\linewidth]{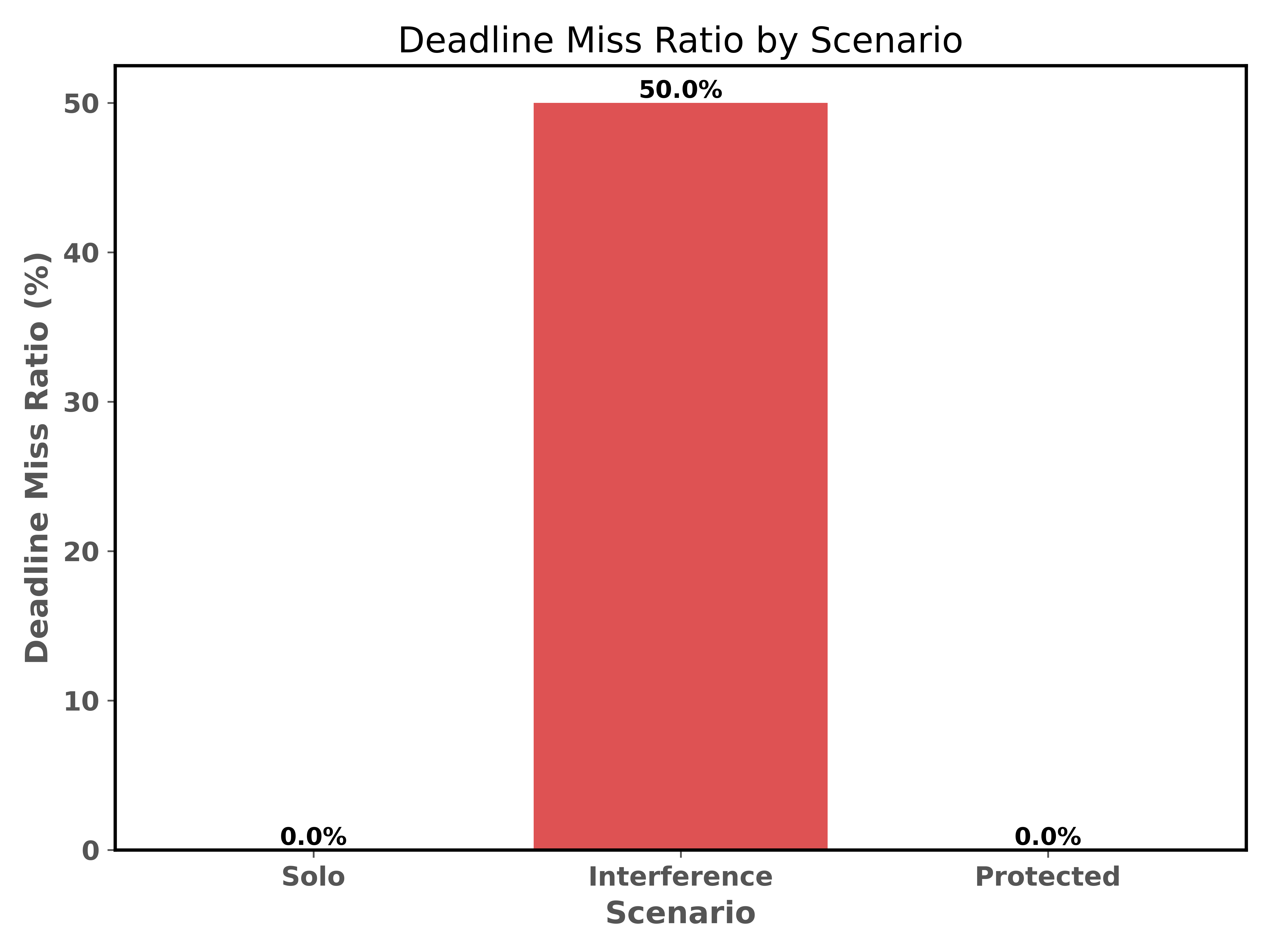}
    \caption{Deadline Miss Ratio comparison.}
    \label{fig:miss_ratio}
\end{figure}

The results of our evaluation are summarized in Figures~\ref{fig:execution_times}, \ref{fig:cdf}, and \ref{fig:miss_ratio}.

Figure~\ref{fig:execution_times} plots the execution time for each of the 100 jobs. In the \textit{Solo} case (green), execution time is stable and well below the 30ms deadline. In the \textit{Interference} scenario (red), memory contention causes severe execution time spikes overlapping every two consecutive jobs, frequently exceeding the deadline by up to 2$\times$. In the \textit{Protected} scenario (blue), the freezer mechanism successfully detects contention and pauses the aggressor.

The Cumulative Distribution Function (CDF) in Figure~\ref{fig:cdf} further illustrates the impact of our approach on the probability of jobs satisfying their performance deadlines. The \textit{Interference} curve indicates that a significant portion of jobs suffer from extreme delays. The \textit{Protected} curve is steep and closely aligns with the \textit{Solo} baseline, demonstrating that our approach effectively eliminates the worst-case outliers caused by memory interference.

Finally, Figure~\ref{fig:miss_ratio} quantifies the number of jobs that our victim task suffers in each execution scenario. 
% The \textit{Solo} scenario results in a no a deadline miss for all the jobs, however the \textit{Interference} execution scenario results in approximately \textbf{50\%}, which is unacceptable for safety-critical systems. In contrast, the \textit{Protected} scenario achieved a \
In the \textit{Solo} execution scenario, $\tau_{victim}$ meets all deadlines as expected. Under \textit{Interference} execution, approximately 50\% of jobs miss their deadline. This ratio is consistent with the period relationship: $\tau_{victim}$ ($T=100$\,ms) and $\tau_{noise}$ ($T=200$\,ms) overlap on every second job of $\tau_{victim}$, causing contention induced slowdown precisely when both tasks execute concurrently. Such a miss rate is unacceptable for safety-critical applications. On the other hand, the \textit{Protected} configuration eliminates all deadline misses by freezing $\tau_{noise}$ during periods of high contention exceeding the allowed threshold, ensuring $\tau_{victim}$ completes within its deadline..

%% file: 2-Related_work.tex
\section{Related work}
\label{sec:related}
The survey by Lugo et al.~\cite{lugo2022survey} categorizes interference mitigation techniques based on the targeted shared resource (e.g. memory, cache, bus) and their integration into the schedulability analysis. Maiza et al.~\cite{maiza2019survey} provide a complementary taxonomy focused on timing verification, distinguishing between isolation-based and contention-aware approaches, including hardware partitioning, scheduling strategies, and hybrid solutions. While prior approaches often target individual aspects of interference, our work introduces a unified methodology that encompasses both hardware and software dimensions to enable integrated analysis and mitigation.

% \subsection{Micro-architectural Modeling}

%\subsection{Interference quantification using machine learning}
Prior work has addressed the challenge of predicting \multicore contention impact on program execution time, primarily by learning a global contention ratio for whole programs. Brando et al.~\cite{brando2022ML} proposed the use of Quantile Regression Neural Networks to estimate the contention-induced slowdown.
%(expressed as the ratio delta = execution time under contention / execution time in isolation). 
Their model leverages event monitors collected during isolated execution to predict this delta value for each program. By tuning the quantile parameter, their model aim to reduce underestimations and provide safer timing budgets during early design phases. Courtaud et al.~\cite{courtaud2019improving} approached the problem from a different angle. They introduced a rich set of microbenchmarks to emulate diverse memory behaviours and revealed the limitations of purely bandwidth-based characterizations. By profiling qualitative and quantitative memory features using Valgrind, they trained a random forest model to predict memory contention overheads. While both works significantly improve the accuracy of contention ratio prediction at the whole-program level, they do not address the localization of contention effects within the code. In contrast, our work formulates the problem as an inverse machine learning task: given global contention observations and static/cache analysis for each instruction, we aim to identify the code regions most susceptible to shared resource contention. This finer granularity enables targeted optimizations and better root-cause analysis of timing variations under interference.

% \subsection{Compilation Methods}

Prior work on dataflow graph fusion like TileFlow~\cite{zheng2023tileflow} and FuseFlow~\cite{lacouture2025fuseflow} has been focused on operator fusion in the context of DNN models deployed on accelerators.
These frameworks are more focused on reducing the amount of memory transfers via fusion, whereas our approach is focused on optimizing the resource usage for more constrained, embedded platforms.
MAESTRO~\cite{kwon2020maestro} describes dataflow through data-centric notations, calculating performance metrics through iteration analysis rather than our model-inference approach.

Other works on on-line interference analyses for multithreaded programs~\cite{zhao2016characterizing} focus on modeling the execution of an application and predicting interference.
This data is then used to influence the scheduling decisions by the hypervisor running all applications.
This is in contrast to our approach that focuses on a deterministic execution model and on-line adjustments of the resources made available by the underlying system. The system described by the authors can explicitly not handle applications with an over-saturation of threads.

Techniques that focus on optimizing the cache layout of multithreaded applications~\cite{sarkar2008compiler} have also shown promising results.
However, they are focused on optimizing cache usage alone, neglecting the overhead introduced by task switching and the changes this brings to the cache layout.

% \subsection{cGroups}

Linux cgroups provide a flexible framework for resource management and have been adapted to address challenges across multiple domains.

In the context of real-time systems, Chen et al. \cite{chen2021schedguard, chen2023schedguard++} 
proposed SchedGuard, a temporal protection framework that uses cgroups to prevent untrusted tasks from executing during specific time segments, protecting against scheduler-based side-channel attacks. Their approach demonstrates the effectiveness of cgroup-based mechanisms for enforcing execution constraints. Similarly, our work exploits cgroups to control task execution, although we focus on mitigating memory contention rather than timing-based security threats. % We use cgroup statistics to detect when monitored task pairs exceed contention thresholds, then employ the freeze mechanism to prevent their concurrent execution in multicore systems.

Andriaccio et al.~\cite{andriaccio2025scheduling} present a cgroup-based real-time scheduler integrated with the Linux deadline server infrastructure to enforce temporal isolation in IoT and edge environments. By associating deadline servers with cgroups, they reserve runtime $Q$ and period $P$ for groups of fixed-priority tasks, supporting multicore migrations to maintain schedulability under multiprocessor models. Validated on FastFlow streaming applications, it achieves lower response-time variance and higher throughput than default schedulers. Unlike their focus on timing predictability for dynamic IoT workloads, our method leverages cgroups for memory-centric isolation, monitoring statistics to trigger freezes and serialize contending tasks, thereby addressing contention vulnerabilities without relying on bandwidth reservations.

For mixed-criticality systems, Kim et al. \cite{kim2018reducing} used cgroups to separate critical and non-critical tasks, then employed the CPUFreq governor to throttle non-critical cgroups when memory 
contention was predicted, indirectly reducing memory bandwidth consumption through CPU frequency scaling. 
In contrast, our work directly prevents co-execution of memory-intensive task  pairs using the cgroup freeze mechanism, avoiding the need for frequency scaling and its associated performance overhead on non-contending tasks.

Beyond the real-time domain, cgroups are extensively applied in cloud computing and virtualized environments, recent work by Volpert et al. \cite{volpert2024hidden, volpert2025detecting} examines noisy-neighbor effects that persist despite CPU isolation using cgroups. They present a workload-agnostic, online detection technique that instruments the Linux scheduler with eBPF to collect runtime metrics—notably scheduling latency and preemption frequency—and use those signals to identify interference between supposedly isolated cgroups. Their results show that scheduler-level interactions can break the practical isolation guarantees of cgroups, motivating adaptive, runtime-aware countermeasures. Building on these findings, our contribution extends cgroups beyond static limits by combining runtime contention monitoring and reactively prevent concurrently running task pairs when measured interference exceeds configurable thresholds.%: we use cgroup statistics to track memory interference and the freezer subsystem to reactively prevent concurrently running task pairs when measured interference exceeds configurable thresholds.

%% file: 6-Conclusion.tex
\section{Conclusion \& Future Work}
\label{sec:conconlusion}

% In this paper, we proposed an interference-aware design methodology for high-performance multi-core systems, our methodology focuses on reducing memory interference through a combination of formal analysis, profiling, and system-level control. These components are integrated in a structured way: low-level interference knowledge feeds both code and deployment decisions, while timing diversity adds a safety mechanism at runtime. For future work, we plan to extend this methodology to integrate control over GPU applications. This extension will address unique interference patterns in heterogeneous CPU-GPU architectures and NUMA architectures. 

In this paper, we presented the initial steps of an interference-aware design methodology for high-performance multi-core systems. The current results are preliminary but they outline a workflow that combines formal analysis, profiling, and system-level control to mitigate memory interference. Future work will investigate the scalability of the approach, its robustness across diverse workloads and hardware platforms, and its applicability to heterogeneous CPU–GPU and NUMA architectures.